\documentclass[RNAAS]{aastex62}

\begin{document}

\title{GRB 210619B: first gamma-ray burst detection by the novel polarimeter MOPTOP}

\correspondingauthor{Manisha Shrestha}
\email{mshrestha1@arizona.edu}

\author[0000-0002-4022-1874]{M. Shrestha}

\affiliation{Astrophysics Research Institute, Liverpool John Moores University, Liverpool Science Park IC2, 146 Brownlow Hill, \\
Liverpool L3 5RF, UK}
\affiliation{Steward Observatory, University of Arizona, 933 North Cherry Avenue, Tucson, AZ 85721-0065, USA}

\author{I. A. Steele}
\affiliation{Astrophysics Research Institute, Liverpool John Moores University, Liverpool Science Park IC2, 146 Brownlow Hill, \\
Liverpool L3 5RF, UK}

\author{S. Kobayashi}
\affiliation{Astrophysics Research Institute, Liverpool John Moores University, Liverpool Science Park IC2, 146 Brownlow Hill, \\
Liverpool L3 5RF, UK}

\author{R. J. Smith}
\affiliation{Astrophysics Research Institute, Liverpool John Moores University, Liverpool Science Park IC2, 146 Brownlow Hill, \\
Liverpool L3 5RF, UK}

\author{H. Jermak}
\affiliation{Astrophysics Research Institute, Liverpool John Moores University, Liverpool Science Park IC2, 146 Brownlow Hill, \\
Liverpool L3 5RF, UK}

\author{A. Piascik}
\affiliation{Astrophysics Research Institute, Liverpool John Moores University, Liverpool Science Park IC2, 146 Brownlow Hill, \\
Liverpool L3 5RF, UK}

\author{C. G. Mundell}
\affiliation{Department of Physics, University of Bath, Claverton Down, Bath, BA2 7AY, UK}


\section{Abstract}
GRB~210619B was a bright long gamma-ray burst (GRB) which was optically followed up by the novel polarimeter MOPTOP on the Liverpool Telescope (LT). This was the first GRB detection by the instrument since it began science observations. MOPTOP started observing the GRB 1388 seconds after the Swift Burst Alert Telescope (BAT) trigger.  The $R$ band light-curve decays following a broken power law with a break time of 2948 s after the trigger. The decay index values are $\alpha_1 = 0.84 \pm 0.03$ (pre-break) and $\alpha_2 = 0.54 \pm 0.02$ (post-break), indicating that the observation was most probably during the forward shock-dominated phase. We find a polarization upper limit of $\sim7$\%. In the forward shock we expect the polarization to mostly come from dust in the local ambient medium which only produces low degrees of polarization.  Hence our non-detection of polarization is as expected for this particular burst.

\section{Introduction}
GRBs are short-lived, very energetic cosmological events. When ejected material in GRB jets interacts with the local ambient medium, afterglow emission is observed at various wavelengths \citep{Piran_1999,Zhang_2004}. Due to their great distance and their orientation, GRB jets are not spatially resolved via traditional astronomical techniques. Instead, we must rely on photometric variability and any polarimetric signal to decipher the structure and magnetic field properties. However, due to the transient nature of GRBs, traditional polarimeters that make sequential measurements of polarization states can induce an artificial polarization signal. With this in mind, a series of polarimeters RINGO (2006 - 2009), RINGO2 (2010 -  2012), and RINGO3 (2013 - 2020) that used a rapidly rotating polaroid analyser \citep{Jermak_2016, Arnold_2017} were developed for the Liverpool Telescope (LT). The newest addition to this series is a novel polarimeter  MOPTOP \citep{Shrestha_2020} which is designed for rapid optical photometric and polarimetric follow-up of GRBs. MOPTOP utilizes a rapidly rotating half-wave plate and a polarization beam splitter to observe the fading GRB. The instrument started science observation in October 2020 and GRB 210619B was the first GRB detected by the instrument. In this research note, we present the light curve and a polarization upper limit for GRB 210619B and provide a brief interpretation of these observations.

\section{Observations}
GRB 210619B was triggered by Swift BAT at 23:59:25 UT \citep{Avanzo_2021}. It has $T_{90} = 54.7$ s \citep{Poolakkil_2021} and was observed at a redshift of 1.937 \citep{Postigo_2021} and \citet{Caballero_2023} reported $E_{\gamma, iso} = 4.05 \times 10^{54}$ erg. The burst was followed up by various ground-based telescopes as reported in GCN circulars \footnote{https://gcn.gsfc.nasa.gov/}. LT started the follow-up $\sim 23$ minutes after the trigger using MOPTOP in the $R$ band.  This was then followed by LT IO:O imager observations  \footnote{https://telescope.livjm.ac.uk/TelInst/Inst/IOO/} in the $r$ band. We note an offset in magnitude between the two instruments ($\sim$ 0.2 mag) due to a slight difference in filter passband.

For both IO:O and MOPTOP, basic CCD reduction with bias subtraction, dark subtraction, flat fielding and World Coordinate System fitting is done internally \footnote{https://telescope.livjm.ac.uk/TelInst/Pipelines/}. The processed images are then used to extract photons counts of the sources using AstroPy {\sc Photutils} package \citep{Bradley_2019}. For MOPTOP and IO:O photometric analysis, we calibrate magnitude using calibration stars from the APASS catalog \citep{Henden_2016}. In the case of MOPTOP we follow the `one camera technique' procedure outlined in \cite{Shrestha_2020} to get polarization values\footnote{Even though the `two camera technique' as shown in \cite{Shrestha_2020} produces higher polarization accuracy, during the observation GRB 210619B there were issues with one of the two cameras. Thus we utilize images from only one camera for our photometry and polarization analysis.}

\section{Discussions}
\begin{figure}[h!]
\begin{center}
\includegraphics[width=\textwidth]{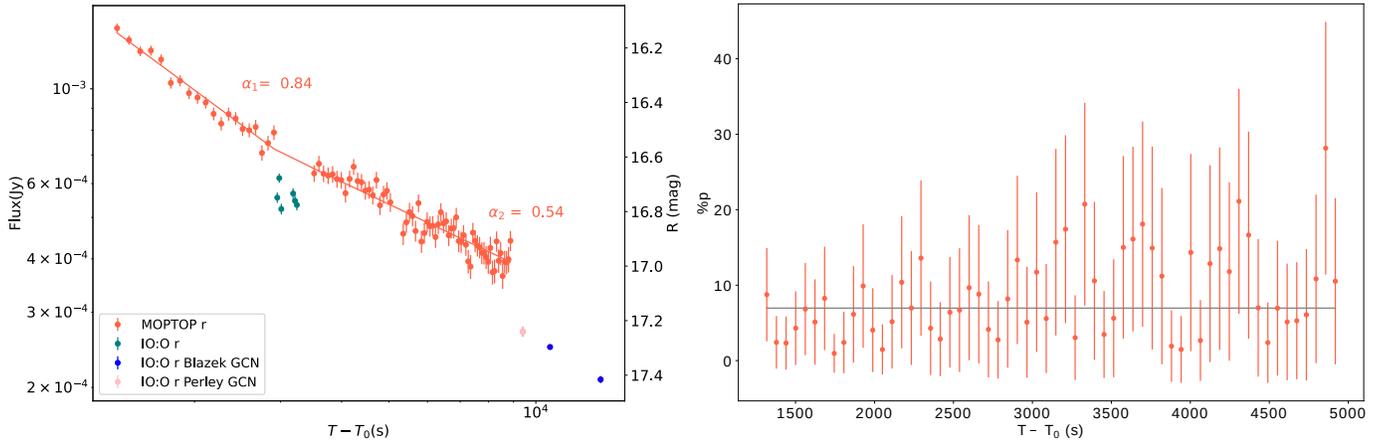}
\caption{(left) Light curve of GRB 210619B with data from the MOPTOP polarimeter in red and IO:O imager in teal and includes three data points from GCN circulars. The best fit broken-power law curve for MOPTOP data is shown in a solid red line with a break time of 2948 s after the trigger and decay index before $\alpha_1 = 0.84  \pm 0.025$ and after $\alpha_2 = 0.54 \pm 0.018$ the break time is shown. Both the y and x-axis are in log scale. (right) Polarization degree with respect to time since the trigger. The gray solid line represents a median value of $6.95 \%$. \label{fig:1}}
\end{center}
\end{figure}

Figure ~\ref{fig:1} left panel presents photometric results from MOPTOP $R$ band observations and IO:O observations along with the last three data points are IO:O $r$ observations from GCN by \cite{Perley_2021,Blazek_2021}. We fit the MOPTOP data both with a single power law and a broken power law (for the case of broken power law we let the break time be a free parameter). We found a broken power law with a break time at $2948 \pm 30$ seconds after the trigger gave the best reduced $\chi^2$ of 1.19 with a degree of freedom 84. From this fit, we get a decay index before the break to be $\alpha_1 = 0.84 \pm 0.025$ and after the break to be $\alpha_2 = 0.54 \pm 0.018$. This shows that we are probably observing the phase where forward shock from the afterglow is dominant \citep{Kobayashi_2000}.

The right panel of Fig.~\ref{fig:1} shows how the polarization degree varies with time. A gray solid line represents median $\% p = 6.95$. For the position of GRB, we used the Galactic extinction value from \cite{Schlegel_1998} $E_{B-V} = 0.1715 \pm 0.005$. We used the relation $p^V \le 9 E_{B-V}$ from \cite{serkowski_1975} to calculate the upper limit in polarization degree in V band and then use relation $p/p_{max} =\exp[-K {\rm ln}^2(\lambda_{max}/\lambda)]$  to estimate galactic interstellar polarization (ISP) in the $R$ band where we assume $\lambda_{max}$ to be V band and $p_{max}$ to be $p^V$ and $K  = -0.10 + 1.86 \lambda_{max}$ given by \cite{Wilking_1982}. Hence we get galactic ISP in the $R$ band to be $1.52\%$ which we don't correct in our results. Though the median value is greater than ISP, the error bars for most of the data crosses 0 which signifies that the observed polarization is upper limit. Thus the non-detection of polarization for the case of GRB~210619B is reasonable as our observation time period is mostly forward shock dominated  as seen in the light curve. \cite{Mandarakas_2023} also made polarimetric observations of the GRB~210619B with $5 \sigma$ detection of $1.5 \%$. Their low polarization values are in good agreement with our observations.

\bibliography{rnaas.bib}
\bibliographystyle{aasjournal.bst}






\end{document}